\documentclass[prb,twocolumn,draft,amsmath,showpacs]{revtex4}
\usepackage{graphics}

\begin{document}

\bibliographystyle{prsty}
\input epsf

\title {Thermal conductivity of single crystalline MgB$_2$}

\author {A.V. Sologubenko, J. Jun, S. M. Kazakov, J. Karpinski, H.R.
Ott}
\affiliation{Laboratorium f\"ur Festk\"orperphysik, ETH H\"onggerberg,
CH-8093 Z\"urich, Switzerland}

\date{\today}

\begin{abstract}
The $ab$-plane thermal conductivity $\kappa$ of single-crystalline 
hexagonal MgB$_2$ has been measured as a function of 
magnetic field $H$ with orientations both parallel and perpendicular to the 
$c$-axis and at temperatures between 0.5 and 300~K. 
In the mixed state, $\kappa(H)$ measured at constant temperatures 
reveals features that are not typical for common type-II 
superconductors.
The observed behavior may be associated with the field-induced 
reduction of two superconducting 
energy gaps, significantly different in magnitude. 
A nonlinear temperature dependence of the electronic thermal conductivity is observed in the field-induced 
normal state at low temperatures. This behavior is at variance with 
the law of 
Wiedemann  and Franz, and suggests an unexpected instability of the 
electronic subsystem in the normal state at $T \approx 1 {\rm ~K}$. 
\end{abstract}
\maketitle

\section{Introduction}

The discovery of superconductivity in MgB$_{2}$ below an 
unexpectedly high
critical temperature $T_c$ of the order of 40~K\cite{Nagamatsu01}
initiated intensive studies of its physical properties. 
Numerous results indicate that the 
superconducting state of  MgB$_{2}$ is conventional in
the sense that the electron pairing is mediated by the
electron-phonon interaction. In most reports the superconducting
energy gap is claimed to be nodeless, compatible with $s$-wave
pairing.
However, various types of experiments,\cite{Bouquet01_b,Wang01,Giubileo01,Giubileo01cm,Szabo01,
Chen01,Laube01,Tsuda01,Junod01cm,Schmidt02,  
Kim02cm,Iavarone02cm} mainly using 
powder or polycrystalline samples and often surface-sensitive, have given 
evidence for two gaps of different magnitude in the quasiparticle excitation spectrum of this 
superconductor. Calculations of the Fermi surface of this material\cite{Kortus01}
reveal three-dimensional sheets ($\pi$-bands) and two-dimensional 
tubes ($\sigma$-bands)
and it seems quite possible that the gap values for these different 
parts of the Fermi surface differ substantially.
It has been argued\cite{Liu01} that
the hole-like quasiparticles on the 2D parts of the
Fermi-surface experience the larger superconducting gap
with a maximum value close to $1.76 k_B T_c$, as predicted by the 
original weak-coupling BCS
theory.\cite{Bardeen57}
The second and smaller gap is associated
with the 3D sheets of Fermi surface. 
This intriguing situation and other possible anomalous 
features of this seemingly simple metallic compound ought to be 
checked experimentally on single crystalline material.

Below we present the results of measurements of the thermal conductivity 
$\kappa$ and the electrical resistivity $\rho$ parallel to the 
basal $ab$-plane of 
the hexagonal crystal lattice of  MgB$_2$ as a function of temperature 
$T$ between 0.5 and 300~K, and varying magnetic fields $H$ between 0 
and 50~kOe, oriented both parallel and
perpendicular to the $c$-axis.
In the mixed state, the observation of a rapid increase of
the electronic thermal conductivity with increasing $H$ at
constant $T \ll T_{c}$ is consistent with  a
field-induced reduction of the smaller of the two superconducting
energy gaps.
Another important observation is that of an
unusual nonlinear temperature dependence of the electronic thermal
conductivity at $T \ll T_{c}$ in the field-induced normal state,
in obvious disagreement with the law of Wiedemann and Franz. 

\section{Sample and experiment}

The investigated single crystal with dimensions of
$0.5\times 0.17\times 0.035$ mm$^{3}$  was grown
with a high-pressure
cubic anvil technique, similar to the one presented
in Ref.~\onlinecite{Lee01} but with a slightly different thermal 
treatment,
described in Ref.~\onlinecite{Angst02}. The high quality of similar crystals from the
same batch was confirmed by single crystal X-ray diffraction and energy dispersive (EDX) X-ray analysis.
A standard uniaxial heat flow method
was used for the $\kappa(T,H)$ measurements.
Temperatures between 0.5 and 2.4~K were achieved in a non-commercial 
$^{3}$He cryostat and the regime between 2 and 300~K was covered by 
using a  $^{4}$He flow cryostat. The thermometers for monitoring the 
temperature difference in the $\kappa(T,H)$ measurements were  Chromel-Au+0.07\%Fe
thermocouples with thin wires (25 $\mu$m diameter) for $T\geq 2 {\rm
~K}$ and a pair of
ruthenium oxide thermometers below 2.4~K. In the region of temperature 
overlap, the results of the measurements of $\kappa(T)$ gave 
identical values for both types of thermometers.
Since the thermopower
of the Au+Fe alloy is strongly field-dependent at low temperatures,
special efforts were made to ensure a reliable calibration of the thermocouples in
magnetic fields.
Possible errors due
to the thermal conductivity of connecting wires are estimated to be below
1\% of the total measured thermal conductivity.
Additional  measurements of the electrical resistivity $\rho(T)$ in the $ab$-plane in magnetic
fields  oriented along the hexagonal $c$-axis were made as
well. The electrical resistivity was measured by employing a 4-contact
configuration with the same contacts used for measuring the voltage and the temperature
difference, respectively.

\section{Results}

\subsection{Electrical resistivity}

\begin{figure}[t]
 \begin{center}
  \leavevmode
  \epsfxsize=1\columnwidth \epsfbox {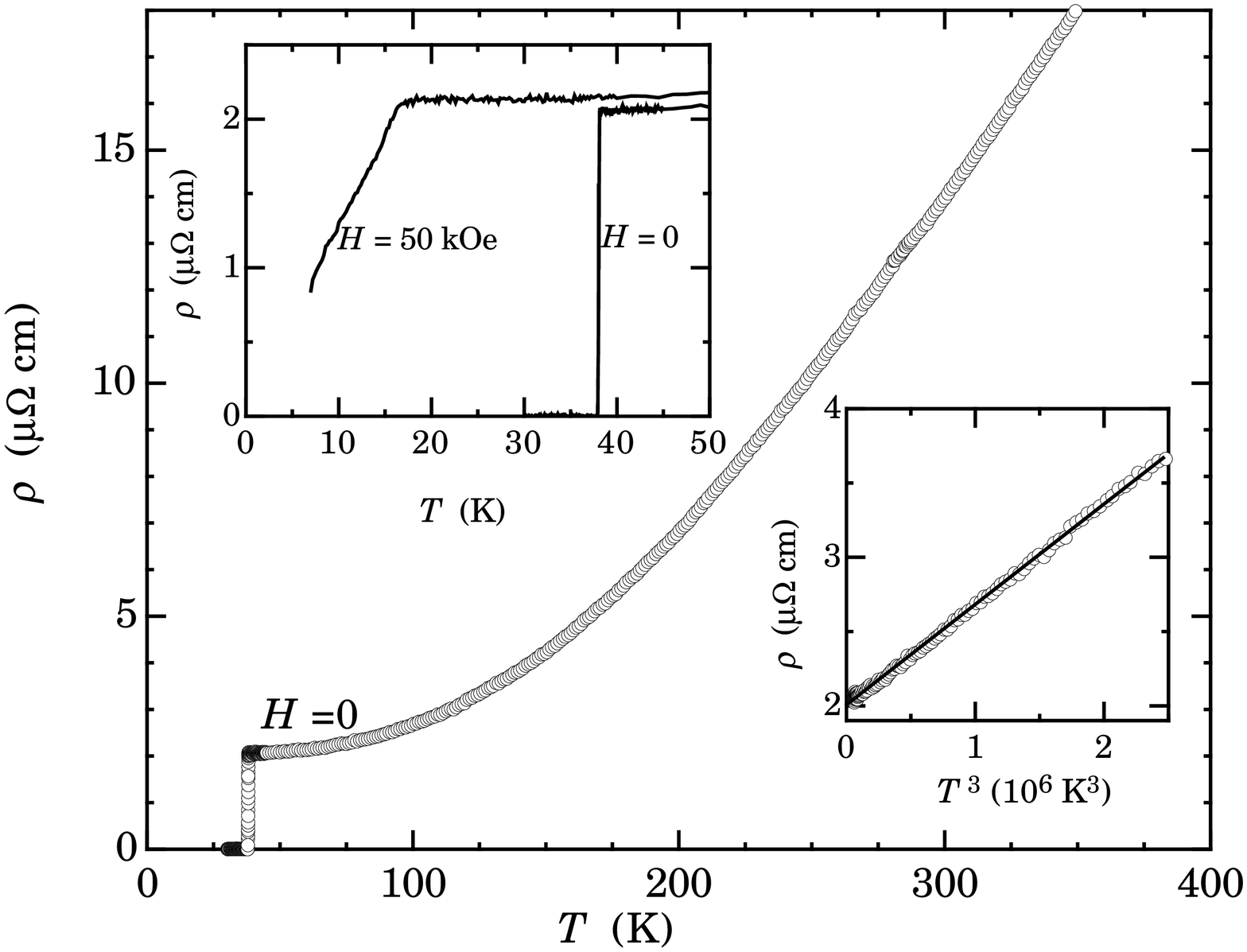}
   \caption{
  In-plane electrical resistivity $\rho(T)$ of hexagonal MgB$_{2}$.
  The upper inset emphasizes the
  low temperature part for $H=0$ and 50~kOe oriented along the $c$-axis. The
  lower inset demonstrates the cubic temperature dependence of $\rho(T)$ 
  for $T \leq 130 {\rm ~K}$.
  }
\label{R}
\end{center}
\end{figure}

The electrical resistivity $\rho(T)$ in zero magnetic field is
presented in Fig.~\ref{R}.
The narrow ($\Delta T_{c}=0.15$~K) superconducting transition
occurs at $T_c=38.1$~K. As it is shown in the lower inset of
Fig.~\ref{R}, in the temperature region between $T_{c}$ and 130~K,
$\rho(T)$ may well be  approximated by $\rho = \rho_{0} + A
T^{3}$, where $\rho_{0}$ and $A$ are constants. At temperatures
below 50~K, $\rho \approx \rho_{0} = 2.0$~$\mu\Omega$~cm. A cubic
$\rho(T)$ dependence has often been  observed in multiband
transition metals and is associated with interband electron-phonon
scattering.\cite{Mott35,Wilson38} The upper inset of Fig.~\ref{R}
emphasizes $\rho(T)$ close to the superconducting transitions in
fields of $H=0$ and $H=50$~kOe, respectively. For our sample, the bulk upper
critical field  for the field direction parallel to the $c$-axis
$H_{c2}^{c}$ is about 30~kOe at zero temperature  and decreases
with increasing temperature,\cite{Sologubenko02} as shown in the 
inset of Fig.~\ref{Tsweeps}. Therefore
the data for $H=50$~kOe were obtained with the bulk of the sample in 
the normal state. The abrupt slope change 
in $\rho(T,H=50~{\rm kOe})$ at 17~K  is most likely related
to the onset of spurious superconductivity in the surface region of 
the sample.\cite{Sologubenko02,Welp02cm}

\subsection{Thermal conductivity}

\begin{figure}[t]
 \begin{center}
  \leavevmode
  \epsfxsize=1\columnwidth \epsfbox {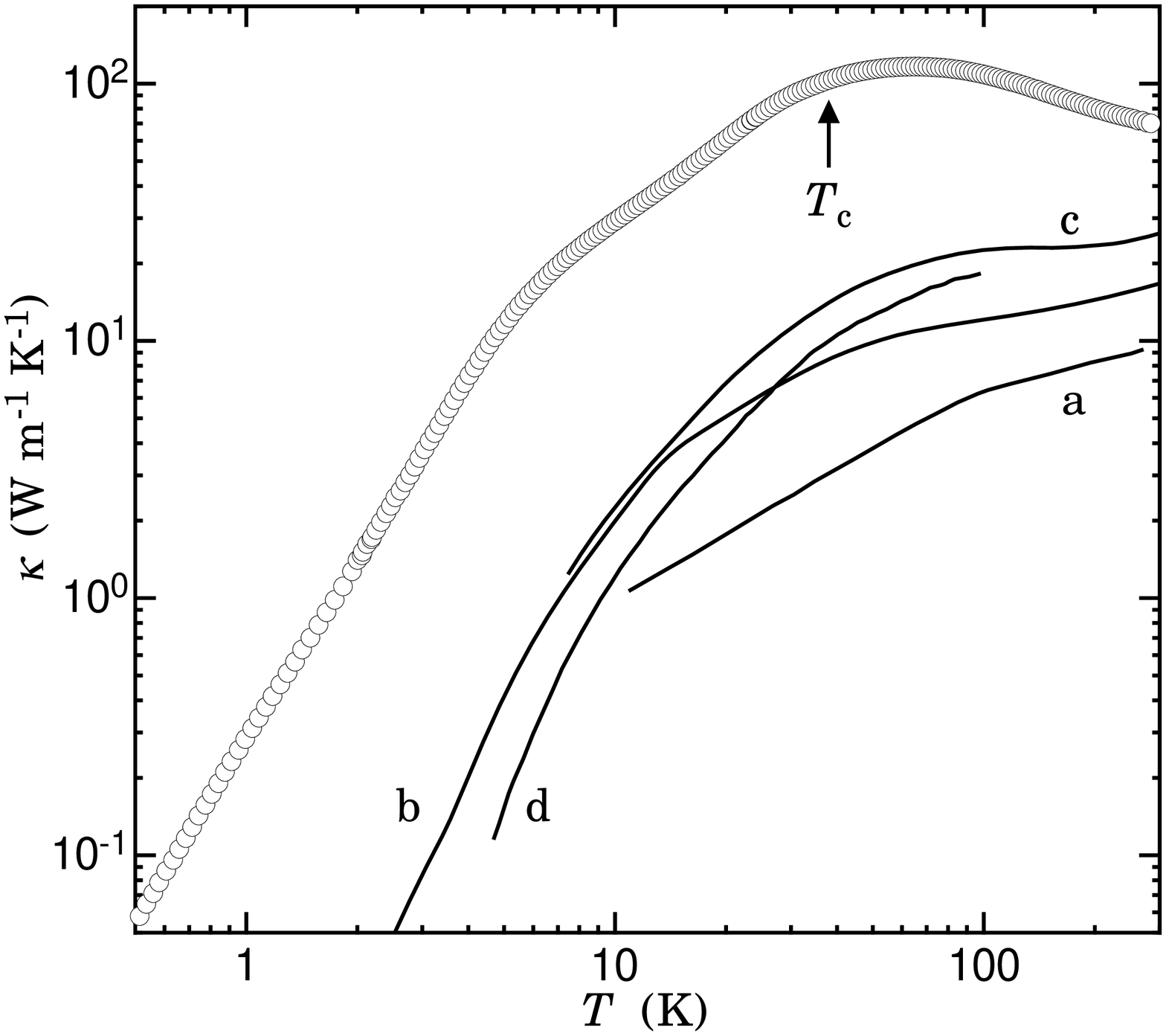}
   \caption{
  Thermal conductivity vs temperature in the $ab$-plane of MgB$_2$
  in zero magnetic field.
  The solid lines represent $\kappa(T)$ measured on
  polycrystalline samples (data from: 
  a - Ref.~\onlinecite{Putti01cm},
  b - Ref.~\onlinecite{Schneider01},
  c - Ref.~\onlinecite{Bauer01},
  d - Ref.~\onlinecite{Muranaka01}).
   }
\label{K}
\end{center}
\end{figure}

The thermal conductivity data $\kappa(T)$ in zero magnetic field
are presented in Fig.~\ref{K}. The $\kappa(T)$ values
are about an order of magnitude higher than previously reported
for polycrystalline samples.\cite{Muranaka01,Bauer01,Putti01cm,Schneider01}
Also the overall temperature dependence of
$\kappa$ is quite different from those earlier data. Instead of a
monotonous increase with temperature we note  a distinct maximum
of $\kappa(T)$ at $T \sim 65$~K. The cause of these differences is
obviously the strong influence of intergrain boundaries on the
heat transport in polycrystals, which masks the intrinsic
mechanisms of quasiparticle scattering. 
As may be seen in Fig.~\ref{K},
no anomaly in $\kappa(T)$ provides evidence for the superconducting transition at
$T_c$.
A distinct change of slope in $\kappa(T)$ at approximately 6~K is 
observed, however.

\begin{figure}[t]
 \begin{center}
  \leavevmode
  \epsfxsize=1\columnwidth \epsfbox {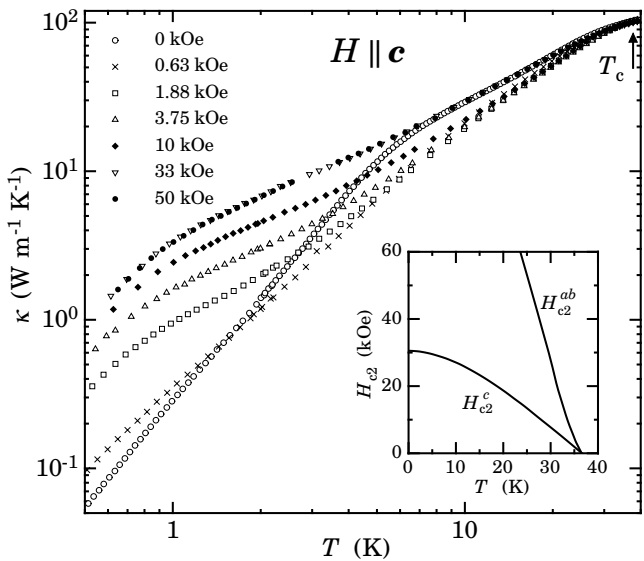}
   \caption{
Thermal conductivity vs temperature in the $ab$-plane of MgB$_2$
  for several values of magnetic fields
  parallel to the $c$-axis.
  The arrow indicates the zero-field critical temperature $T_{c}$.
  The temperature dependences of the upper critical fields
  $H_{c2}^{c}$ and $H_{c2}^{ab}$ as established in
  Ref.~\onlinecite{Sologubenko02} are shown in the inset.
  }
\label{Tsweeps}
\end{center}
\end{figure}

\begin{figure}[t]
 \begin{center}
  \leavevmode
  \epsfxsize=1\columnwidth \epsfbox {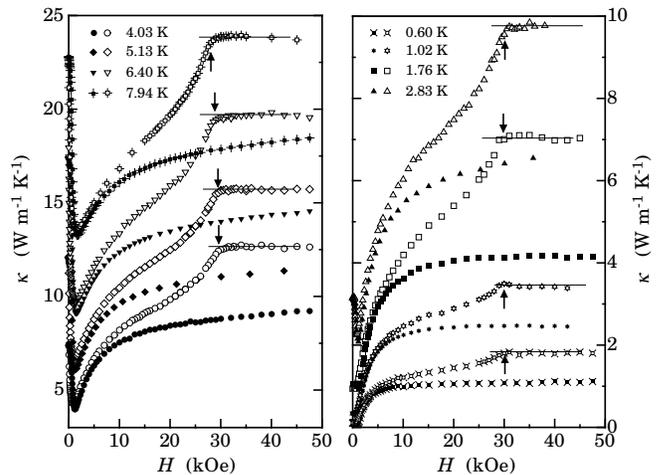}
   \caption{
  Thermal conductivity in the basal plane of MgB$_{2}$ vs $H$ at several fixed temperatures.
  The arrows denote the upper critical field
  $H_{c2}$ for $H \parallel c$. The closed and open  symbols correspond to the field
  direction perpendicular and parallel to the $c-$axis, respectively.
  }
\label{Hsweeps}
\end{center}
\end{figure}

In Fig.~\ref{Tsweeps} we display $\kappa(T)$ in the range between 0.5 and
40~K, measured at selected fixed magnetic fields, oriented parallel to the
$c$-axis.
The practically
overlapping curves for $H = 33$ and 50~kOe 
indicate that for this field orientation and these $H$ values, the
normal state has been reached  in the whole covered temperature range.
The initial decrease of $\kappa$ with increasing field is relatively large at higher
temperatures, but is progressively reduced with decreasing
temperatures. It finally turns into an increase of $\kappa(H)$ at 
constant temperatures below approximately 1~K.
This behavior of the low-temperature thermal conductivity in the mixed state $0 < H <
H_{c2}$ is better illustrated in Fig.~\ref{Hsweeps} where
we present the  $\kappa(H)$ curves, measured at constant temperatures
below 8~K for field directions both parallel and perpendicular to
the $c$-axis. 
The typical features of these curves are the rapid initial decrease 
of $\kappa$ with increasing field, narrow minima in $\kappa(H)$ at 
field values that are low with respect to $H_{c2}$, and a 
subsequent $s$-shape type increase of $\kappa$ with further increasing 
field. The latter feature is particularly pronounced for the
$\kappa(H)$ curves at the lowest temperatures.
The low-field increase is practically independent of the field direction
as may clearly be seen in Fig.~\ref{Hsweeps}, where data for $H \parallel
c$ (open symbols) and $H \perp c$ (solid symbols) are displayed for 
comparison. The increasing slope at higher fields and $H \parallel c$  
is certainly caused by approaching the normal state at $H_{c2}$. This 
trend is not observed 
for $H\perp c$, for which $H_{c2}$ is estimated to be about
130~kOe at low temperatures, obviously far beyond our experimental 
possibilities.

\section{Discussion}
The thermal conduction of a superconductor is usually provided by
electronic quasiparticles ($\kappa_e$) and phonons ($\kappa_{\rm ph}$), such that
\begin{equation}\label{eKappa}
\kappa = \kappa_e + \kappa_{\rm ph}.
\end{equation}
Upon decreasing the temperature to below $T_{c}$ in zero magnetic
field, the reduction of the number of unpaired electrons leads to
a decrease of $\kappa_{e}$ and an increasing  $\kappa_{\rm ph}$.
The overall behavior of $\kappa(T)$ in the superconducting state
depends on the relative magnitudes of $\kappa_e$ and $\kappa_{\rm
ph}$ and also on the strength of the electron-phonon interaction.
Applying external magnetic fields induces vortices in the sample.
The quasiparticles associated with the vortices not only provide
additional contributions to phonon scattering and hence  a
reduction of $\kappa_{\rm ph}$, but also enhance $\kappa_{e}$. The
competition of these two processes leads to $\kappa(H)$ curves as
shown in Fig.~\ref{Hsweeps}.

In what follows, we first discuss the low-temperature temperature
part of $\kappa(T)$ in the field-induced normal state, i. e., for $H \geq 
H_{c2}$. Subsequently we discuss $\kappa(T)$ in  
the superconducting state for $H=0$ and finally turn to the data of 
$\kappa(H)$, measured at constant temperatures in the mixed state, i. e., 
$H < H_{c2}$.

\subsection{Thermal conductivity in the normal state}
One of the main problems encountered in the analysis of the thermal 
conductivity of a conductor is that 
a separate identification of the two terms in Eq.~(\ref{eKappa}) at
arbitrary temperature is not straightforward. In the normal state,
a convenient and often used way to estimate $\kappa_e$ is to
employ the Wiedemann-Franz law (WFL), relating the electrical
resistivity and the electronic contribution to the thermal
conductivity via
\begin{equation}\label{eWFL}
\kappa_{e}(T) = L_{0} T / \rho(T),
\end{equation}
where $L_{0}=2.45\times 10^{-8}$ W~$\Omega$ K$^{-2}$ is the Lorenz
number. The validity of this law requires an elastic scattering of
electrons and it is well established that Eq.~(\ref{eWFL}) is
applicable if the scattering of electrons by defects dominates.
This is usually true at  low temperatures, where $\rho(T) \approx
\rho_{0}$.  In our case, the data in Fig.~\ref{R} suggest that
Eq.~(\ref{eWFL}) is applicable at temperatures below about 50~K. 
Inserting the appropriate numbers into 
Eq.~(\ref{eWFL}) suggests that, at temperatures $T_{c} \leq T \leq
50$~K, $\kappa_{e}(T)$ provides about half of the total thermal
conductivity. As we demonstrate below, the applicability of the
WFL for MgB$_2$ at low temperatures is
questionable, however.

Before discussing the validity of WFL at low temperatures, we note
the complications that arise from unusual features in the 
$\rho(T)$ data shown in Fig.~\ref{R}, most likely caused by the influence of 
superconductivity in 
minor regions of the sample with
enhanced $T_{c}$ and $H_{c2}$, different from the corresponding bulk 
values.
The onset of this superconductivity is
clearly seen in the upper inset of Fig.~\ref{R} by inspecting $\rho(T)$ 
measured
in a field of 50~kOe, substantially exceeding $H_{c2}(0)$, the maximum  bulk upper critical
field of the same sample.\cite{Sologubenko02} 
Superconducting traces 
in a minute
fraction of the sample may cause a considerable reduction of the
total measured electrical resistivity but leave the
thermal conductivity virtually unchanged. 
In this case a failure of 
the WFL would not be surprising.  
However, because the electrical resistivity
of the sample above $T_{c}$ is practically
temperature-independent, we can, with a great deal of certainty,
expect that the intrinsic electrical resistivity  of {\em the bulk}
remains constant also at lower temperatures. 
The constant residual resistivity
$\rho_{0}$ is caused by defects which actually set the maximum
mean free path for electrons, independent of temperature. 
Indeed, results by Xu {\em et al.}
\cite{Xu01} of $\rho(T)$ measurements on single crystalline 
MgB$_{2}$ at magnetic fields which presumably exceed the
upper critical field of the minor phase, demonstrate that 
in, e.g., $H=90 {\rm ~kOe}$ the electrical resistivity remains
constant with decreasing temperature down to at least 2~K.

\begin{figure}[t]
 \begin{center}
  \leavevmode
  \epsfxsize=1\columnwidth \epsfbox {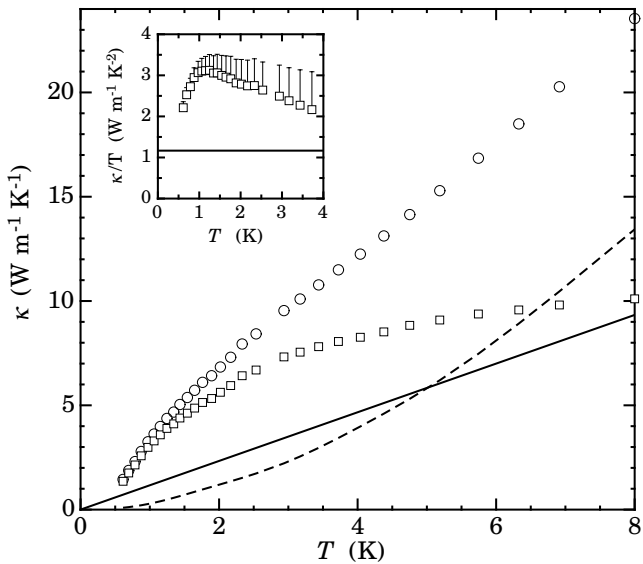}
   \caption{
  Normal-state thermal conductivity measured in $H= 33 {\rm ~kOe}$ 
  between 0.6 and 8~K 
  (open circles). The upper limit of the phonon thermal 
  conductivity $\kappa_{\rm ph}^{\rm max}$ (dashed line) and the lower limit of the electronic 
  contribution $\kappa_{e}^{\rm min}$ (open squares) are calculated as described in the text.
  The solid line represents the electronic contribution 
  $\kappa_{e}^{\rm WFL}$ calculated 
  using the  Wiedemann-Franz law in Eq.~(\ref{eWFL}).
  In the inset, we plot  $\kappa_{e}^{\rm min}$ (open squares)  and 
  $L_{0}/\rho_{0}$ (solid line), respectively. The error bars in the 
  inset mark the maximum uncertainty of $\kappa_{e}/T$.
  }
\label{Ke_est}
\end{center}
\end{figure}

In Fig.~\ref{Ke_est}, we plot as a solid line, the normal-state
electronic thermal conductivity $\kappa_{e}^{\rm WFL}$ below 8~K, calculated using
Eq.~(\ref{eWFL}) with the experimental value $\rho_0 = 2.1 {\rm ~\mu \Omega cm}$ for $H=33 {\rm ~kOe}$.
The measured total thermal conductivity, shown by open circles, is
considerably higher than $\kappa_{e}^{\rm WFL}$ across the entire covered 
temperature range. 
For the estimate of the upper limit of the phonon contribution 
we assume that the minima of $\kappa(H)$ 
shown in Fig.~\ref{Hsweeps} are caused by the competition of a decreasing $\kappa_{\rm ph}$ 
and an increasing $\kappa_{e}$. 
With this interpretation it is clear that the values of 
$\kappa_{\rm min}(H)$ represent at most the maximum value of the 
lattice contribution $\kappa_{\rm ph}$.
The smooth interpolation between these minimum values 
of $\kappa(H)$ for different temperatures, denoted as 
$\kappa_{\rm ph}^{\rm max}$, is shown as the broken 
line in Fig.~\ref{Ke_est}. 
The difference between the measured 
thermal conductivity at $H = 33 {\rm ~kOe}$ and  the upper limit of the phonon contribution,
$\kappa_{e}^{\rm min} = \kappa - \kappa_{\rm ph}^{\rm max}$, obviously representing 
the lower limit of the electronic contribution, is shown in 
Fig.~\ref{Ke_est} by open squares. 
It may be seen that at least below 8~K, the electronic contribution 
is considerably larger than predicted by WFL.
With increasing temperature, $\kappa_{e}^{\rm min}$ approaches the
WFL prediction, as is demonstrated in the inset of 
Fig.~\ref{Ke_est}. 
Nevertheless, it is impossible to identify  the
temperature limit where the validity of the WFL is recovered if such 
a limit exists at all. 
Considering our procedure it is clear that the true
electronic contribution exceeds our estimate,
particularly towards the upper end of the considered temperature range.
However, at very low temperatures where $\kappa(H=0) \ll \kappa_{n}$
(e. g.,  at 0.60~K the zero field thermal conductivity is less than  6\% 
of the normal-state thermal conductivity),
$\kappa_{e}^{\rm min}$ must be very close to the true $\kappa_{e}$. 
Thus the observed deviation from the WFL
and in particular its temperature dependence, revealing a 
peak-like structure 
of $\kappa_{e}/T$ vs $T$, shown in the 
inset of Fig.~\ref{Ke_est}, is a reliable result of our investigation. 

The violation of the Wiedemann-Franz law at low temperatures is 
very unusual because 
the validity of this law is expected to hold for 
the Fermi-liquid ground state of common metals.
A similar set of data of Hill an coworkers for (Pr,Ce)$_2$CuO$_4$ was interpreted 
as evidence for a 
breakdown of the Fermi-liquid theory for this oxide material.\cite{Hill01} 
The non-Fermi-liquid behavior of cuprates was speculated to be 
the consequence of 
possible spin-charge separation, a scenario that is considered in 
relation with 
high-$T_{c}$ superconductivity.\cite{Anderson87}  
The same arguments are hardly relevant for the case of MgB$_{2}$ 
where a spin-charge separation is not expected. 
Another explanation for the results of 
Ref.~\onlinecite{Hill01} 
was recently offered in Ref.~\onlinecite{Si01}.
It was argued that a peak-type structure of $\kappa_{e}/T$ plotted 
versus $T$ might be the consequence of some kind of transition 
leading to a gap formation in the electronic excitation spectrum of 
the normal state. This scenario would lead to a peak of $\kappa_{e}/T$ 
at $T_{0}$, close to the transition temperature, an exponential decrease of $\kappa_{e}/T$ 
below $T_{0}$, and a gradual approach to the  Wiedemann-Franz value of 
$L_{0}/\rho_{0}$ at $T>T_{0}$. 
Such a peak at $T_{0} \sim 1 {\rm ~K}$ indeed follows from our 
analysis of the  MgB$_{2}$ data (see the inset of Fig.~\ref{Ke_est}). 
Unfortunately our data set does not extend to low enough temperatures 
in order to confirm the exponential temperature dependence of $\kappa_{e}$ 
well below $T_{0}$.

\subsection{Thermal conductivity in the superconducting state 
($H=0$)}\label{SectionH0}

A rather unexpected feature in the temperature dependence of $\kappa$ 
in zero magnetic field is 
the absence of even the slightest manifestation of the transition at  $T_{c} = 
38.1 {\rm~K}$,
as has already been mentioned in previous reports on $\kappa(T)$ for polycrystalline 
materials.\cite{Muranaka01,Bauer01,Putti01cm,Schneider01}
This observation is quite unusual for superconductors with non-negligible phonon heat transport, because 
the opening of the superconducting gap rapidly reduces 
the rate of phonon 
scattering on electrons and should lead to a fast increase of $\kappa_{\rm ph}$ below 
$T_{c}$.  
The assumptions that either the phonon-electron scattering is much 
weaker than phonon-defect scattering, or 
that $\kappa_{\rm ph}$ is negligibly small in the 
vicinity of $T_{c}$,
which, in principle, might explain the absence of a $\kappa(T)$ feature at 
$T_{c}$, are all incompatible with the observation that applying a relatively weak 
external magnetic field of 0.63~kOe, introducing  
some additional quasiparticles in the cores of vortices,  
considerably reduces the thermal 
conductivity at intermediate temperatures (see Fig.~\ref{Tsweeps}). 
The possibility that the enhancement of $\kappa_{\rm 
ph}$ below $T_{c}$ is exactly compensated by a reduction of 
$\kappa_{e}$, is not considered because, for a BCS superconductor,  the 
slope change in  $\kappa_{\rm ph}(T)$  at $T_{c}$ is much more 
pronounced than that in  $\kappa_{e}(T)$.\cite{Bardeen59} 
The latter statement is not true for special cases of extremely clean samples of 
strong-coupling superconductors, such as Pb and 
Hg,\cite{Watson63,Ambegaokar64} where 
the  scattering of electrons by defects is negligibly 
small. Our sample of MgB$_{2}$ with $\rho(300)/\rho(0) \approx 
6.8$ cannot really be regarded as fulfilling the extreme clean 
limit criteria. 
It is possible, however, to account for both the absence of a feature 
in  
$\kappa(T)$ at $T_{c}$ and the obvious slope change in $\kappa(T)$ 
centered around 6~K, by postulating that the 
superconducting energy gap $\Delta(T)$ for quasiparticles which strongly 
interact  
with low-frequency phonons, is considerably smaller than the values
given by the BCS theory.

In the simplest approximation, the phonon thermal conductivity can be 
calculated as
\begin{equation}
\kappa_{\rm ph} = (v^{2}/3) \int C(\omega) \tau(\omega) d\omega, 
\end{equation}
where 
$v$ is the mean sound velocity, and 
$C(\omega)$ and $\tau(\omega)$ are the specific heat and the average 
relaxation time of a phonon mode with frequency $\omega$, respectively. 
The total phonon relaxation rate may be calculated by assuming that 
the simultaneous influence of all independent 
phonon scattering mechanisms $\tau_{i}$ can be accumulated in the form
\begin{equation}\label{eTotal}
\tau^{-1} = \sum_{i} \tau_{i}^{-1}.
\end{equation}
The phonon-electron relaxation time $\tau_{p-e}$ changes most rapidly 
upon the opening of the  superconducting gap. 
In Ref.~\onlinecite{Bardeen59}, the phonon-electron scattering time
in the superconducting state  takes the form
\begin{equation}
\tau_{p-e}^{s}(\omega) = g(\omega/T,\Delta/T) 
\tau_{p-e}^{n}(\omega), 
\end{equation}
where $\tau_{p-e}^{n}(\omega)$ is the 
normal-state relaxation time. 
The function $g(\omega/T,\Delta/T)$ is quite complicated, but its main feature  
is a step-like increase of the phonon relaxation rate at the 
phonon frequency $ \omega = 2 \Delta / \hbar$, a consequence of the 
fact that  a phonon with an energy less than $2 \Delta$ cannot brake a Cooper 
pair and interacts only with quasiparticles that have already been excited above the 
gap.\cite{GeilikmanBook} 
In the ``dominant phonon'' approximation\cite{BermanBook} it is assumed that, at 
temperature $T$, the 
main contribution to the heat transport in the lattice is due to phonons with 
frequencies close to $\omega_{\rm dom}$, where $ \hbar \omega_{\rm 
dom} \approx 3.8 k_{B} T$. For weak-coupling BCS
superconductors, where $\Delta(0)=1.76 k_{B} T_{c}$ and $\Delta(T)$ 
is a standard function tabulated in Ref.~\onlinecite{Muehlschlegel59}, $\hbar 
\omega_{\rm dom}(T) = 2 \Delta(T)$ is always fulfilled at the temperature $0.73 T_{c}$, i. 
e., not far below $T_{c}$. 
This is 
the reason why $\kappa_{\rm ph}(T)$ increases rapidly below $T_{c}$, 
leading to a maximum of the measured $\kappa$, typically close to $T_{c}/2$.\cite{GeilikmanBook}  
For MgB$_{2}$, instead of a peak near $T_{c}/2$, 
$\kappa(T)$ exhibits a distinct feature at about $T_{c}/6 \approx 6 {\rm ~K}$, which can also be 
regarded as a peak-type structure on top of the background which 
decreases  monotonously 
with decreasing $T$. 
This suggests that the relevant superconducting energy gap is equal to $\hbar \omega_{\rm 
dom}$ at much lower temperatures than $0.73 T_{c}$. Hence,   
for the quasiparticles which scatter phonons  
most effectively, the energy gap $\Delta(T)$ is about 3 times smaller than the values given 
by the original BCS theory.  
 
In principle, more information could have been extracted from our data by 
direct comparison with existing theories for the thermal conductivity in 
multiband superconductors.\cite{GeilikmanBook,Tang71,Tang72,Kumar72} 
However, since the WFL seems to be invalid in the normal state of MgB$_{2}$ 
at low temperatures, any attempt to analyze $\kappa(T)$ quantitatively in the 
superconducting state 
is hampered by the difficulties in separating $\kappa_{e}$ and 
$\kappa_{\rm ph}$ in a reliable manner.

\subsection{Thermal conductivity in the mixed state}
Before discussing the features of $\kappa$ in magnetic fields, 
it is important to note that the zero-field values of $\kappa$ 
at temperatures $T \ll T_{c}$ are almost entirely due to 
the phonon contribution. 
The electronic thermal conductivity in the superconducting state $\kappa_{e,s}(T)$  
can be estimated using the theory of 
Bardeen, Rickayzen, and Tewordt \cite{Bardeen59}. In their model
\begin{equation}\label{eKe}
\kappa_{e,s} = \kappa_{e,n} f(y),
\end{equation}
where
\begin{eqnarray}\label{eKes}
f(y) &=& {\frac{  2F_1(-y) + 2y\ln(1+e^{-y}) + \frac {y^2} {1+e^{y}} }   { 
2F_1(0)  } }, \quad(T<T_{c})\nonumber\\
{\rm and}&&\\
f(y) &=& 1,\quad(T \geq T_{c}),\nonumber
\end{eqnarray}
as well as $F_{n}(-y) = \int_0^\infty z^{n} (1+e^{z+y})^{-1} dz$, $y=\Delta(T)/k_{B} T$. 
Using for $\kappa_{e,n}$ the values of $\kappa_{e,n}^{\rm min}$ shown in Fig.~\ref{Ke_est}, 
and taking into account the lowest previously claimed value of 
$\Delta(0)=1.7$~meV, \cite{Tsuda01} it may be shown that 
$\kappa_{e,s}$ is negligibly small below 4~K, and therefore, $\kappa(H=0) \approx \kappa_{\rm ph}$.

As it may clearly be seen in Fig.~\ref{Hsweeps}, the $\kappa(H)$ curves 
reveal very similar features at all temperatures below 8~K.
The most intriguing aspect  of these curves is the very 
rapid increase of $\kappa$ at relatively weak fields, after 
the initial decrease of $\kappa_{\rm ph}$.  
This increase is undoubtedly due to a field induced enhancement of the 
number of electronic quasiparticles and thus to an increase of $\kappa_{e}$.
For common type II superconductors in their mixed state, the features of 
$\kappa_{e}(H)$ are expected to depend on the ratio between the electron mean free path $\ell$ 
and the coherence length $\xi_{0}$. 
A rough estimate of $\ell$ can be obtained from the residual 
resistivity $\rho_{0}$ employing the Drude relation 
$\rho_{0} = 3/ [N_{0} \ell v_{F} e^{2}]$, where $N_0$ is the 
electronic density of 
states at the Fermi level and $v_F/3$ is the average in-plane component of 
the Fermi velocity. Using the 
values $N_{0}=0.7 {\rm ~states/(eV~unit~cell)}$ (Ref.~\onlinecite{Kortus01}) and 
$v_{F}=4.9\times 10^{7} {\rm ~cm/sec}$ (Ref.~\onlinecite{Bouquet01_b}), we 
obtain $\ell \approx 80 {\rm ~nm}$. Since this value is considerably larger 
than the in-plane coherence length $\xi_{ab,0}=11.8 {\rm ~nm}$ 
(Ref.~\onlinecite{Sologubenko02}), our sample is in the moderately clean limit. 
In this limit, 
$\kappa_{e}$ is expected to be small at all fields below $H_{c2}$, except close to  
$H_{c2}$, where it grows according to 
\begin{equation}\label{eClean}
\kappa_{e} = \kappa_{e,n} (1-C_{T}(H_{c2}-H)^{1/2}),
\end{equation}
where $\kappa_{e,n} \propto T$ is the normal-state 
electronic thermal conductivity above $H_{c2}$ and $C_{T}$ is a 
temperature-dependent coefficient.\cite{Maki67}
As may be seen in Fig.~\ref{Hsweeps}, for MgB$_{2}$, the largest 
positive slope $d\kappa/dH$ is 
observed well below $H_{c2}$ for both orientations of the magnetic field. 
Although the form of Eq.(\ref{eClean}) implies that $\kappa(H)$  
should scale with the value of $H_{c2}$,   
the thermal conductivity depends only weakly on the field direction up to approximately 6~kOe.
This is amazing because 
the anisotropy of the upper critical field $H_{c2}^{ab}/H_{c2}^{c} \approx 4.2$ at low 
temperatures.\cite{Sologubenko02}

For $H \perp c$, after a steep initial increase, $\kappa(H)$  
reaches a region where it exhibits only a relatively weak $H$-dependence. 
The same tendency is also observed for $H||c$ but it is partly masked 
by yet another increase of $\kappa(H)$ close to $H_{c2}$.
For $H \perp c$ and $T \leq 8$~K, $H^{ab}_{c2}\sim 130$~kOe 
(Ref.~\onlinecite{Sologubenko02} ) 
and 
is not accessible in our experimental setup. Therefore the region of 
weak $H$-dependence extends to the highest fields reached in this 
study.

The field dependence of $\kappa_{e}$, although very different from 
what one would expect from a conventional superconductor,
can qualitatively be explained in terms of a two-band model 
with two energy gaps of different magnitude associated with each band.
Nakai and co-authors\cite{Nakai02} analyzed such a model where one band 
with strong pairing ($L$-band) is responsible for superconductivity, and 
superconductivity in the second band ($S$-band) is induced by  Cooper pair 
tunneling. Consequently, the two bands are   
characterized by a smaller gap $\Delta_{S}$ and a larger gap $\Delta_{L}$, 
and normal-state electronic densities 
of states at the Fermi level $N_{0,S}$ and $N_{0,L}$, respectively. 
The analysis presented in Ref.~\onlinecite{Nakai02} shows that the 
quasiparticle states in the vortices are highly confined 
in the $L$-band but only loosely bound in the $S$-band. Therefore the 
quasiparticle states of the
vortices in the $S$-band start to overlap already in weak fields and 
the resulting density of states 
equals that of the normal-state $N_{0,S}$ at $H \ll H_{c2}$.
The situation can be visualized as a vortex lattice involving the 
$L$-band states, coexisting with  the normal state in the $S$-band where the energy gap is 
suppressed. This model explains very well the behavior of the electronic specific 
heat in a magnetic field.\cite{Yang01,Junod01cm} The field-induced 
suppression of the smaller gap is claimed to be consistent 
with the results of point-contact measurements\cite{Szabo01,Laube01} and recent 
scanning tunneling spectroscopy experiments.\cite{Eskildsen02up}  
In terms of the two-gap model, the saturation of the thermal conductivity 
much below $H_{c2}$ may be regarded as the result of the closing of the energy gap in 
the $S$-band. The heat transport via quasiparticles of the band associated with the larger 
gap is significant only in the vicinity of 
$H_{c2}$, and, eventually, above $H_{c2}$ the full normal state
electronic heat 
transport is restored.
The lack of a substantial 
dependence of $\kappa_{e}$ on the field orientation for $H \ll H_{c2}$ is an  
obvious consequence of the weakly anisotropic 3D nature of the $\pi$-bands. 
From this we conclude that the smaller gap must open in the $\pi$-band. 
The rapid increase of the number of quasiparticles in the $\pi$-bands 
also
naturally explains  the very fast drop of $\kappa_{\rm ph}(H)$ in
small fields, 
because the corresponding excited quasiparticles are the dominant 
scattering centers for phonons at low temperatures.

\begin{figure}[t]
 \begin{center}
  \leavevmode
  \epsfxsize=0.75\columnwidth \epsfbox {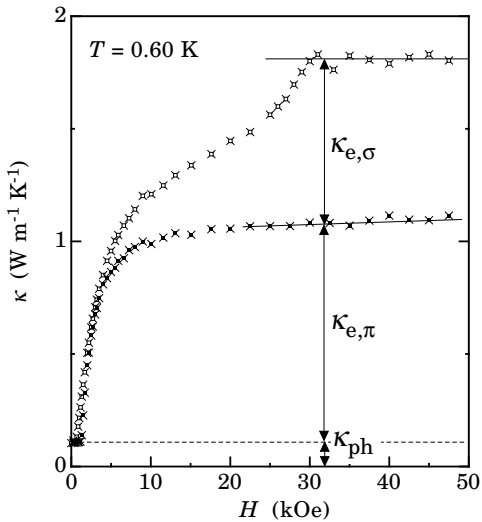}
   \caption{
  Separation of the individual contributions of the $\sigma$- and 
  $\pi$-band quasiparticles, 
  and the phonons to the normal state thermal conductivity of MgB$_{2}$ 
  at $T = 0.60 {\rm ~K}$. 
  }
\label{K060}
\end{center}
\end{figure}

A more quantitative analysis of $\kappa_{e}(H)$ can be made for the lowest temperatures 
where, as may be seen from Fig.~\ref{Tsweeps}, the phonon contribution is relatively small in comparison with 
the field-induced electronic contribution. 
At very low temperatures, the phonon scattering by 
electronic quasiparticles is less effective,\cite{BermanBook} 
therefore  $\kappa_{\rm ph}(H>0)$ should not much deviate  from 
$\kappa_{\rm ph}(H=0)$.
Indeed, at temperatures of 0.60 and 1.02~K, there 
is no initial decrease  of $\kappa(H)$ in small fields. 
Assuming that the phonon contribution is essentially $H$-independent and that the 
smaller gap is completely suppressed in fields exceeding 20~kOe, we 
may establish the individual contributions to $\kappa_{e}$ of the quasiparticles 
associated with either the
$\sigma$- or the $\pi$-band. This is illustrated in Fig.~\ref{K060}.
A possible reduction of $\kappa_{\rm ph}$ with increasing $H$ could 
slightly change this ratio in favor of $\kappa_{e, \pi}$, but only 
by a few percent. 
The ratio $\kappa_{e,\pi}/\kappa_{e,\sigma}$ is 0.57/0.43, as estimated 
from the 
$\kappa(H)$ data at $T=0.60 {\rm ~K}$. This ratio is remarkably close 
to the ratio of the densities of electronic states in the two bands 
$N_{0,\pi}/N_{0,\sigma}$ of 0.58/0.42, as calculated by Liu {\em et 
al.}\cite{Liu01} and 0.55/0.45  by Belashchenko {\em 
et al.}\cite{Belashchenko01}. Similar ratios  of 
0.55/0.45 and 0.50/0.50 have been extracted from  
tunneling spectroscopy measurements\cite{Eskildsen02up} and 
from specific heat experiments,\cite{Junod01cm}  respectively. 
From this comparison, we reach the important conclusion 
that the 
electron mean free paths on different sheets 
of the Fermi surface are close to being equal. Indeed, the electronic thermal 
conductivity may be 
calculated from $\kappa_{e} = C_{e} v_{F} 
\ell/3$, where $C_e$  is the electronic specific 
heat. Since $C_{e,i} \propto N_{0,i}$, and the $ab$-components of the 
Fermi velocity
$v_{F,i}$ ($i=\pi,\sigma$) are similar for different sheets of the Fermi surface in 
MgB$_{2}$,\cite{Antropov01cm} the equality 
$\kappa_{e,\pi}/\kappa_{e,\sigma}\approx N_{0,\pi}/N_{0,\sigma}$ is 
tantamount to saying 
that, at low temperatures, the ratio  of the electron mean free paths in different bands 
$\ell_{\pi}/\ell_{\sigma}$ is close to unity.  
This observation is  essential in view of the current 
discussion of the possibly different impurity scattering rates 
in different bands of electronic states of MgB$_{2}$.\cite{Kuzmenko02,Mazin02cm,Mazin02cm_b}

Although the absence of 
any particular feature at $T_{c}$ in the zero-field $\kappa(T)$ data, discussed in 
Section \ref{SectionH0},  gives only 
qualitative support for the existence of parts of the Fermi surface 
with a gap much smaller than predicted by the standard BCS theory,  
the magnetic field induced variation of the low-temperature thermal conductivity 
may be regarded as strong evidence in favor of the multigap scenario. 

\begin{figure}[t]
 \begin{center}
  \leavevmode
  \epsfxsize=1\columnwidth \epsfbox {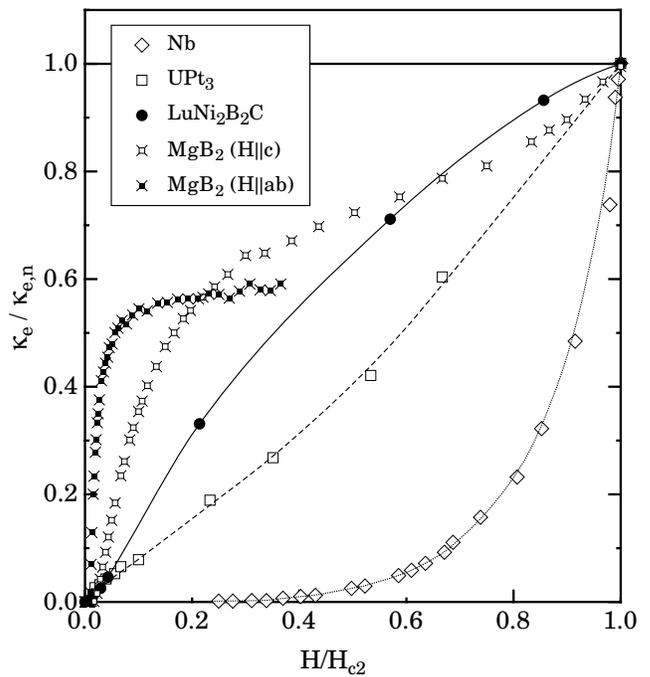}
   \caption{
  The electronic thermal conductivity normalized to its normal state 
  value vs. $H/H_{c2}$. The data for MgB$_{2}$ are 
  from this work, the results for Nb, UPt$_{3}$, and LuNi$_{2}$B$_{2}$C are from 
  Refs.~\onlinecite{Lowell70}, \onlinecite{Suderow97}, and \onlinecite{Boaknin01}, 
  respectively.
  }
\label{BS}
\end{center}
\end{figure}

At the same time we believe that the model of Haas and Maki\cite{Haas02} 
for explaining thermodynamic and optical properties of MgB$_{2}$ is 
not appropriate. They  proposed the ${\bf 
k}$-dependence of a single energy gap to adopt the form of a prolate ellipsoid. 
Our claim is based on the comparison of our data set 
with similar results 
for materials with strongly anisotropic gap functions. In 
Fig.~\ref{BS}, we redraw a figure from 
Ref.~\onlinecite{Boaknin01}, which compares 
the field-induced variation of 
$\kappa_{e}$ at temperatures well below $T_{c}$  for different 
conventional and  
unconventional superconductors, amended by our data for MgB$_{2}$ 
at $T=0.60{\rm ~K}$. 
The Nb data\cite{Lowell70} reveal the typical salient features of  a clean, 
almost isotropic $s$-wave 
superconductor, 
and confirms the very weak energy 
transport by quasiparticles far below $H_{c2}$, in agreement with 
Eq.~(\ref{eClean}). A considerably faster, 
almost linear in $H$, increase of $\kappa_{e}$ is observed for a  
superconductor with nodes in $\Delta({\bf k})$, 
here exemplified  by UPt$_{3}$.\cite{Suderow97} 
A similar $H$-variation of $\kappa_{e}$ has been observed for 
LuNi$_{2}$B$_{2}$C, which led the authors of Ref.~\onlinecite{Boaknin01} 
to claim an anisotropy of the energy gap $\Delta_{\rm max}/\Delta_{\rm 
min} > 10$. The increase of $\kappa_{e}(H)$ 
in MgB$_{2}$ is much faster than in any of these materials.
For the field direction $H \perp c$, more than half of the normal-state thermal 
conductivity is restored already at $H = 0.05 H_{c2}$! This means that, 
with increasing $H$, 
instead of the gradual increase of the number of quasiparticles contributing 
to the heat transport, that is characteristic of single-gap anisotropic 
superconductors, we are dealing with the suppression of 
an energy gap on a significant portion of the total Fermi surface  by 
relatively  weak magnetic
fields. Thus the features of $\kappa_{e}(H)$ of MgB$_{2}$ 
displayed in Fig.~\ref{BS} may be regarded as a natural 
consequence of the existence of two different gaps.

\section{Summary}

Our $\kappa(T,H)$ data provide evidence for a rapid 
field-induced enhancement of quasiparticles in the superconducting 
state of MgB$_{2}$ well below $H_{c2}$, consistent with an efficient 
field-induced closing of the smaller energy gap, 
thus provoking a 
fast growth of the electronic thermal conductivity. 
At higher fields, the growth of $\kappa_{e}(H)$ tends to 
saturate until,  
in the vicinity of $H_{c2}$, the contribution to $\kappa$ 
from the electrons of the band associated with the larger energy gap grows 
rapidly, merging  into the practically field-independent thermal 
conductivity in the normal state above $H_{c2}$. 

At low temperatures, the electronic thermal conductivity of the 
field-induced normal state  is nonlinear in $T$
and deviates considerably from the prediction of the Wiedemann-Franz law. 
This deviation peaks at about 1~K, suggesting the existence of some  
transition  provoking a gap formation in the electronic excitation 
spectrum close to this temperature.

\acknowledgments
We acknowledge useful discussions with I.L. Landau, R.
Monnier, M. Chiao,  and M. Sigrist.
This work was financially supported in part by
the Schweizerische Nationalfonds zur F\"{o}rderung der Wissenschaftlichen
Forschung.

\end{document}